\documentclass[aps,prl,superscriptaddress,reprint,showpacs]{revtex4-1}
\usepackage{amssymb}
\usepackage{amsmath}
\usepackage{amsfonts}
\usepackage{graphicx}
\usepackage[colorlinks=true,linkcolor=blue,urlcolor=blue,citecolor=blue]{hyperref}
\usepackage[all]{hypcap}
\usepackage {xcolor}

\begin{document}

\title{Unconventional temperature dependence of the anomalous Hall effect in HgCr$_2$Se$_4$}

\author{Shuai~Yang}
\affiliation{Beijing National Laboratory for Condensed Matter Physics, Institute of Physics, Chinese Academy of Sciences, Beijing 100190, China}
\affiliation{School of Physical Sciences, University of Chinese Academy of Sciences, Beijing 100049, China}
\author{Zhilin~Li}
\affiliation{State Key Laboratory for Artificial Microstructure and Mesoscopic Physics, Beijing Key Laboratory of Quantum Devices, Peking University, Beijing 100871, China}
\affiliation{Beijing National Laboratory for Condensed Matter Physics, Institute of Physics, Chinese Academy of Sciences, Beijing 100190, China}
\author{Chaojing~Lin}
\author{Changjiang~Yi}
\author{Youguo~Shi}
\affiliation{Beijing National Laboratory for Condensed Matter Physics, Institute of Physics, Chinese Academy of Sciences, Beijing 100190, China}
\affiliation{School of Physical Sciences, University of Chinese Academy of Sciences, Beijing 100049, China}
\author{Dimitrie~Culcer}
\affiliation{School of Physics, University of New South Wales, Sydney 2052, Australia}
\affiliation{ARC Centre of Excellence in Future Low-Energy Electronics Technologies, UNSW Node, Sydney 2052, Australia}
\author{Yongqing~Li}
\email{yqli@iphy.ac.cn}
\affiliation{Beijing National Laboratory for Condensed Matter Physics, Institute of Physics, Chinese Academy of Sciences, Beijing 100190, China}
\affiliation{School of Physical Sciences, University of Chinese Academy of Sciences, Beijing 100049, China}
\affiliation{Songshan Lake Materials Laboratory, Dongguan, Guangdong 523808, China}

\date{\today}

\begin{abstract}
 At sufficiently low temperatures, many quantum effects, such as weak localization, electron-electron interaction (EEI), and Kondo screening, can lead to pronounced corrections to the semiclassical electron transport. Although low temperature corrections to longitudinal resistivity, ordinary Hall resistivity, or anomalous Hall resistivity are often observed, the corrections to three of them have never been simultaneously detected in a single sample. Here, we report on the observation of  $\sqrt T$-type temperature dependences of the longitudinal, ordinary Hall and AH resistivities at temperatures down to at least 20 mK in $n$-type HgCr$_2$Se$_4$, a half-metallic ferromagnetic semiconductor that can reach extremely low carrier densities. For the samples with moderate disorder, the longitudinal and ordinary Hall conductivities can be satisfactorily described by the EEI theory developed by Altshuler \textit{et al.}, whereas the large corrections to  AH conductivity are inconsistent with the existing theory, which predicts vanishing and finite corrections to AH conductivity for EEI and weak localization, respectively.
\end{abstract}


\maketitle

A key issue of condensed matter physics is to study the effects of electron phase coherence, many-body interaction, disorder, and the interplay between them. At low temperatures, they are manifested in a variety of transport phenomena beyond the semiclassical regime, such as Aharonov-Bohm oscillations~\cite{Webb1985}, weak localization~\cite{Dries1981}, metal-insulator transition~\cite{Kravchenko1994}, fractional quantum Hall effect~\cite{Tsui1982}, unconventional superconductivity~\cite{Steglich1979}, and Kondo effect~\cite{Kondo1964}.
While understanding electron transport in the systems with strong EEIs still poses a great challenge~\cite{Spivak2010}, it is generally accepted that the knowledge of quantum transport in weakly interacting non-magnetic systems is much more complete, thanks to  a series of seminal works during late 1970s to early 1980s~\cite{Abrahams1979,Gorkov1979,Altshuler1980,Hikami1980,Altshuler1983}.  Deviations of electron transport from the semiclassical behavior, often referred to as low temperature anomalies or quantum corrections, have been observed in numberous experiments and can be described satisfactorily with the theories taking account of the weak localization (antilocalization) and EEI effects~\cite{Bergmann1984,Lee1985,Altshuler1985}. In contrast, much less attention has been devoted to low temperature corrections to electron transport  in magnetic systems.

In the systems with broken time-reversal symmetry, low temperature corrections to the  anomalous Hall effect (AHE) render a valuable and unique probe to study the interplay between disorder, phase coherence, and EEI, since the AHE is governed by microscopic processes that are different from those for longitudinal conductivity and ordinary Hall (OH) effect~\cite{Nagaosa2010,Xiao2010}.
The first work in this direction was reported in 1991 by Bergmann and Ye, who found that the temperature dependence of the anomalous Hall (AH) conductivity is nearly zero in ultrathin Fe films despite the fact that both the longitudinal and the AH resistances exhibit $\ln{T}$-type dependences at low temperatures~\cite{Bergmann1991}. The absence of low temperature correction was explained by W\"{o}lfle and coworkers, who suggested theoretically that the EEI correction to certain contributions to the AH conductivity vanishes for arbitrary strengths of impurity scattering for the systems with mirror symmetry~\cite{Langenfeld1991, Muttalib2007}. Although it was later observed that the $\ln{T}$-type of corrections to AH conductivity can become substantial in ferromagnetic (FM) metal thin films with sufficiently strong disorder, such corrections were nevertheless attributed to the weak localization effect~\cite{Mitra2007, Lu2013, Ding2015, Wu2016}. To the best of our knowledge, only a few groups have studied the theory of \textit{low temperature} corrections to AHE and none of them has proposed a mechanism other than the weak localization~\cite{Langenfeld1991,Dugaev2001,Muttalib2007}.


In this paper, we report a surprising observation of a large correction to the AH conductivity in HgCr$_2$Se$_4$, a low electron density FM system in which the weak localization effect can be excluded by the temperature dependence of longitudinal conductivity, and by the nearly vanishing temperature dependence of the OH conductivity. This contrasts with previous studies of the low temperature corrections to AHE, in which the measurements were carried out on FM metals~\cite{Mitra2007,Xiong2010,Lu2013} or heavily doped magnetic semiconductors~\cite{Mitra2010}. The high carrier densities in those systems make it very difficult to measure the correction to OH effect reliably. In HgCr$_2$Se$_4$ a $\sqrt T$-type of temperature dependence has been observed in all three resistivity components, namely the longitudinal, OH, and AH resistivities. While temperature dependences of the longitudinal and OH conductivities (resistivities) are in accordance with the EEI theory of Altshuler \textit{et al.}, the large $\sqrt T$-type correction to the AH conductivity cannot be explained with existing theory. 

\begin{figure}[htb]
	\centering
	\includegraphics[width=8.5cm]{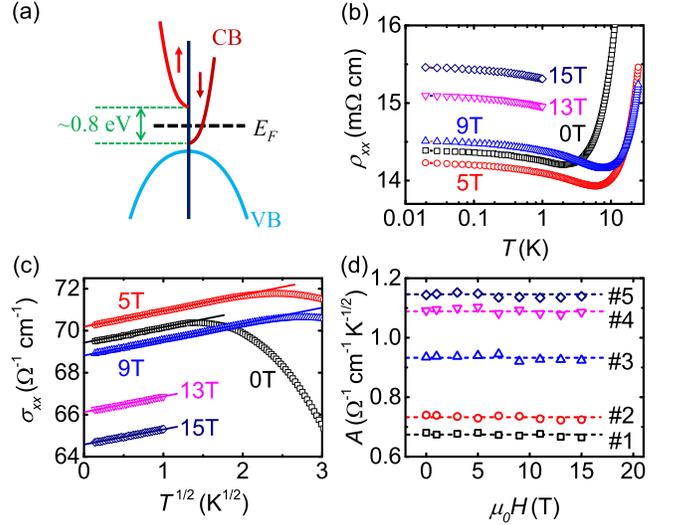}
	\caption{(a) Band diagram of $n$-type HgCr$_2$Se$_4$ in the FM phase. (b) Temperature dependence of $\rho_{xx}$ at $T<30$ K in constant magnetic fields. (c) Longitudinal conductivity $\sigma_{xx}$ plotted as a function of $T^{1/2}$. The data in panels (b) and (c) were taken from sample \#2. (d) Magnetic field dependences of $d\sigma_{xx}/d\sqrt{T}$ of samples \#1-\#5 (See Table~\ref{tabI} \& Eq.~\ref{eq1}).}
	\label{fig1}
\end{figure}

Our measurements were performed on a set of $n$-type HgCr$_2$Se$_4$ single crystals, which have a spinel structure (space group Fd3m) and ferromagnetic ground state with Curie temperature $T_C \approx 105$ K~\cite{Baltzer1966,Lin2016}. At liquid helium temperatures, the electron density $n$ is in a range of 10$^{15}$-10$^{18}$ cm$^{-3}$, several orders of magnitude lower than the density of Cr$^{3+}$ ions. The magnetic properties are mainly determined by superexchange interactions between Cr$^{3+}$ ions~\cite{Baltzer1966,Yaresko2008}, and barely influenced by free carriers~\cite{Lin2016}. At low temperatures, the FM exchange interaction causes a large spin splitting in the conduction band ($\sim$ 0.8 eV), making the $n$-type HgCr$_2$Se$_4$ a half-metallic semiconductor with a narrow band gap~\cite{Arai1973, Selmi1987, Yaresko2008, Guo2012, Guan2015, Note1}, as illustrated in the band diagram in Fig.~\ref{fig1}(a). 
Since the n-type HgCr$_2$Se$_4$ samples can remain metallic down to very low electron densities, they offer a unique magnetic system to study the low temperature corrections caused by the EEI effect. The interaction parameter, defined as $r_s=a/a_B^{*}\propto E_C/E_F\propto n^{-1/3}$ (where $a$ is the Wigner-Seitz radius, $a_B^{*}$ is the effective Bohr radius, $n$ is the carrier density, $E_C$ is the Coulomb energy and $E_F$ the Fermi energy)~\cite{SM}, can be tuned by nearly one order of magnitude in HgCr$_2$Se$_4$. 
In this work, we focus on the samples with $n>1\times 10^{17}$ cm$^{-3}$ so that the transport is not complicated by strong interaction effects ~\cite{Spivak2010,SM}. Given in Table \ref{tabI} are the basic characteristics of five samples (\#1-\#5) on which transport measurements were performed in detail.

\begin{table*}[htb]
\centering
\caption{Basic characteristics of the HgCr$_2$Se$_4$ samples, including carrier density $n$, mobility $\mu$, conductivity $\sigma_0$, resistivity $\rho_0$, disorder parameter $k_F l_e$, diffusion coefficient $D$, and the slope of $\sigma_{xx}$ vs. $\sqrt{T}$ curve (experiment and theory), as well as the corresponding ratios between the low temperature corrections to the AH and longitudinal conductivities (defined as $r_\sigma=\delta \sigma_\mathrm{AH}^\mathrm{N}(T)/\delta\sigma_{xx}^\mathrm{N}(T)$). The values of $n$, $\mu$, $\sigma_0$, $k_F l_e$ and $D$ are evaluated with $\rho_{xx}$  and $R_H$ at $T=0$, which are obtained by extrapolation of the zero-field resistivity and the Hall effect data at finite temperatures to $T=0$.}
\label{tabI}
\begin{ruledtabular}
\begin{tabular}{cccccccccc}
	Sample & $n$ & $\mu$ & $\sigma_0$ & $\rho_0$ & $k_F l_e$ & $D$ & $d\sigma_{xx}/d\sqrt{T}$(exp) & $d\sigma_{xx}/d\sqrt{T}$(theory) & $\left|r_\sigma\right|$ \\
	No. & ($10^{18}$ cm$^{-3}$) & ($\rm cm^2/Vs$) & ($\rm (\Omega\cdot cm)^{-1}$) & ($\rm m\Omega\cdot cm$) & &
	($\rm cm^2/s$) & ($\rm (\Omega\cdot cm)^{-1}K^{-1/2}$) & ($\rm (\Omega\cdot cm)^{-1}K^{-1/2}$)\\
	\hline
	\#1 & 0.85 & 897 & 122    &8.2  & 8.04 & 20.7  & 0.68 & 0.55 & 90 \\
	\#2 & 0.58 & 749 & 69.6  &14.4  & 5.21 & 13.4  & 0.74 & 0.68 & 95 \\
	\#3 & 0.31 & 704 & 34.9  &28.7  & 3.22 & 8.29  & 0.93 & 0.85 & 81 \\
	\#4 & 0.15 & 642 & 15.4  &64.9  & 1.81 & 4.66  & 1.09 & 1.12 & -  \\
	\#5 & 0.11 & 514 & 9.0   &111   & 1.18 & 3.03  & 1.14 & 1.37 & -  \\
\end{tabular}
\end{ruledtabular}
\end{table*}

Fig.~\ref{fig1}(b) shows the temperature dependences of longitudinal resistivity of sample \#2 in various magnetic fields. As the temperature is lowered, the decrease in $\rho_{xx}$  can be attributed to reduced scatterings from spin waves. Further lowering $T$ to liquid helium temperatures leads to a resistivity upturn, which is more pronounced for the samples with higher resistivities~\cite{SM}. The resistivity upturn is insensitive to the external magnetic field: its shape remains nearly unchanged up to at least $\mu_0 H=15$ T. Fig.~\ref{fig1}(c) shows that longitudinal conductivity $\sigma_{xx}$ has a linear dependence on $\sqrt{T}$ down to at least 20 mK, without any sign of saturation as $T$ approaches to zero. More strikingly, the slope of low temperature part of $\sigma_{xx}$-$\sqrt{T}$ curve is nearly constant as magnetic field $\mu_0 H$ is varied from 0 to 15\,T. This feature also appears in all other samples, as shown in Fig.~\ref{fig1}(d).

The $\sqrt{T}$-type correction to $\sigma_{xx}$ has been observed previously in 3D non-magnetic semiconductors and amorphous alloys~\cite{Thomas1982,Sahnoune1992,Neumaier2009}. It can be explained with the theory of EEI in three dimensions (3D)~\cite{Altshuler1985,Lee1985}. The correction to longitudinal conductivity, $\Delta\sigma_{xx}$, is proportional to $\left(\frac 43-\frac 32\tilde F_\sigma\right)\sqrt{\frac TD}$, where the first term in the bracket originates from the exchange interaction, the second one is the Hartree term, $\tilde F_\sigma$ is the screening constant, and $D$ is the diffusion coefficient. For a half metal, in which the electron spins are fully polarized at the Fermi level, the Hartree term is reduced to $\frac 12\tilde F_\sigma$, and the EEI correction to $\sigma_{xx}$ becomes
\begin{equation}
\label{eq1}
\Delta\sigma_{xx}(T)=\frac{e^2}{\hbar}\frac{1}{4\pi^2}\frac{1.3}{\sqrt2}\left(\frac 43-\frac 12\tilde F_\sigma\right)\sqrt{\frac{k_B T}{D\hbar}}=A\sqrt{T}.
\end{equation}

It should be noted, however, that weak localization can lead to additional correction to conductivity at low temperatures~\cite{Lee1985,Bergmann1984}. In a 3D system, the correction also follows a power law: $\Delta\sigma_{xx}(T)\propto{T^{p/2}}$ , in which the value of exponent $p$ is dependent on the dephasing mechanism of electrons. For instance, $p=$ 3/2, 2, and 3 for the Coulomb interactions in the dirty limit and clean limit, and electron-phonon scattering, respectively~\cite{Lee1985}. The linear dependence of $\sigma_{xx}$ on $\sqrt T$ would correspond to $p=1$, which is too small to be attributed to the weak localization effect even if one considers the dephasing caused by spin wave excitations in HgCr$_2$Se$_4$~\cite{Takane2003,SM}. The irrelevance of weak localization is further confirmed by the nearly constant values of $d\sigma_{xx}/d\sqrt{T}$ in magnetic fields up to $\mu_0 H=15$ T, since conductivity correction caused by the weak localization is usually suppressed in a strong magnetic field (on the order of Tesla or less)~\cite{Bergmann1984}. Moreover, the $d\sigma_{xx}/d\sqrt{T}$ values are in good agreement with those calculated with Eq.~\eqref{eq1} and experimental values of the diffusion coefficient~(see Table~\ref{tabI} \& discussions in~\cite{SM}).

\begin{figure}[htb]
	\centering
	\includegraphics[width=8.5cm]{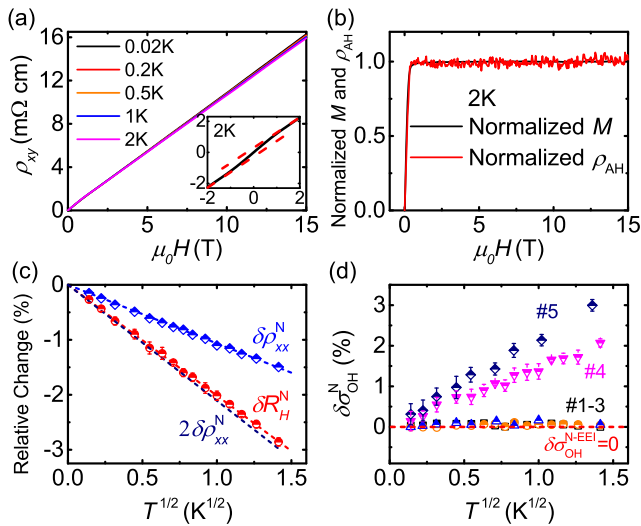}
	\caption{(a) Magnetic field dependence of ordinary Hall (OH) resistivity $\rho_{xy}$ at several temperatures. The inset is a close-up view of the small low-field  curvature manifesting the AHE. (b) Normalized AH resistivity and magnetization plotted as a function of $\mu_0 H$ at $T=2$ K. (c) Relative changes in the OH coefficient ($R_H$) (circles) and $\rho_{xx}$ (diamonds). The dashed line corresponds to $2\delta\rho_{xx}^{\rm N}$. The data in panels (a)-(c) were taken from sample \#2.  (d) $T$-dependences of relative changes in the OH conductivities of samples~\#1-\#5, where $\delta\sigma_\mathrm{OH}^{\mathrm{N}}=[\sigma_{xy}^{\mathrm{OH}}(T)- \sigma_{xy}^{\mathrm{OH}}(0)]/\sigma_{xy}^{\mathrm{OH}}(0)$.}
	\label{fig2}
\end{figure}
Low temperature corrections to the Hall effect have also been observed in HgCr$_2$Se$_4$.
As shown in Fig.~\ref{fig2}(a), the Hall resistivity curves resemble those of a typical FM system with weak AHE. The good linearity in a wide range of magnetic fields ($\sim$2-15\,T) allows for a reliable separation of the AH component from the OH resistivity~\cite{SM}. 
Fig.~\ref{fig2}(a \& b) shows that the normalized AH resistivity, obtained by subtracting the OH contributions from the external magnetic field (denoted as $H$ in this work) and the internal field due to the magnetization of the sample($H_\mathrm{in}$)~\cite{SM}, overlaps very well with the normalized bulk magnetization. 
The obtained Hall coefficient, $R_H$, exhibits a linear dependence on $\sqrt T$. According to the EEI theory~\cite{Altshuler1980,Altshuler1980a}, the relative change in $R_H$, defined as
$\delta R_H^{\rm N}={[R_H(T)-R_H(0)]}/{R_H(0)}={\Delta R_H(T)}/{R_H^0}$, has a magnitude two times the relative change in $\rho_{xx}$, namely $\delta R_H^{\rm N}=2\delta \rho_{xx}^{\rm N}$, where $\delta \rho_{xx}^{\rm N}={[\rho_{xx}(T)-\rho_{xx}(0)]}/{\rho_{xx}(0)}=\Delta\rho_{xx}/\rho_0$. As depicted in Fig.~\ref{fig2}(c), this is nearly perfectly borne out in our experiment. In contrast, the weak localization effect is not expected to modify the OH resistivity but can influence the OH conductivity~\cite{Altshuler1980a}. Fig.~\ref{fig2}(d) shows the nearly vanishing temperature dependence of the OH conductivity $\sigma_{xy}^\mathrm{OH}$ in samples \#1-\#3, further supporting a dominant role of EEI and negligible weak localization effect in these samples. This statement, however, only holds for the weakly disordered (diffusive) transport regime ($k_F l_e\gg$ 1, where $k_F$ is the Fermi wave vector, and $l_e$ is electron mean free path). The finite $T$-dependences of $\sigma_{xy}^\mathrm{OH}$ observed in samples~\#4 and \#5 displayed in Fig.~\ref{fig2}(d) can be attributed to the stronger disorder~\cite{Uren1980}, as evidenced by $k_F l_e$ values approaching one (See Table~\ref{tabI}).

\begin{figure}[htb]
	\centering
	\includegraphics[width=8.5cm]{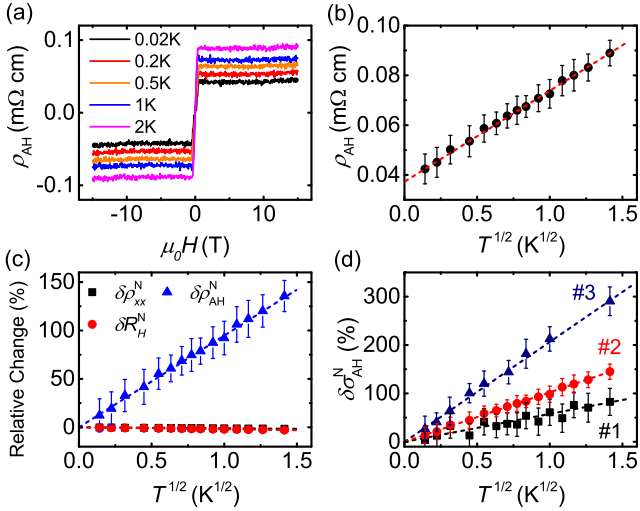}
	\caption{(a) Magnetic field dependence of anomalous Hall (AH) resistivity $\rho_\mathrm{AH}$ at $T=$ 20 mK-2 K. (b) $\sqrt{T}$-dependence of  $\rho_\mathrm{AH}$. (c) $\sqrt{T}$-dependences of relative change in $\rho_{xx}$, $R_H$ and $\rho_\mathrm{AH}$. (d) $\sqrt{T}$-dependence of relative change in the AH conductivity $\sigma_\mathrm{AH}$. The data in panels (a)-(c) were taken from sample \#2, and panel (d) from samples \#1-\#3.}
	\label{fig3}
\end{figure}

Fig.~\ref{fig3} shows the main experimental result of this work. For the samples with moderate disorder (samples~\#1-\#3, with $k_F l_e\approx 3.2$-8.0), both AH resistivity $\rho_\mathrm{AH}$ and AH conductivity $\sigma_\mathrm{AH}$ exhibit a strong linear dependence on $\sqrt T$ at low temperatures(Fig.~\ref{fig3}(a \& b)). As illustrated in Fig.~\ref{fig3}(c), the relative change in $\rho_\mathrm{AH}(T)$, defined as $\delta\rho_\mathrm{AH}^{\rm N}(T)={[\rho_\mathrm{AH}(T)-\rho_\mathrm{AH}(0)]}/{\rho_\mathrm{AH}(0)}=\Delta\rho_\mathrm{AH}/\rho_\mathrm{AH}^0$, reaches nearly 135\% at $T=2$ K in sample~\#2. It is nearly two orders of magnitude larger than the relative changes in longitudinal and OH resistivities, which are only 1.45\% and 2.86\%, respectively, for the same temperature interval (i.e.~0-2 K). The AH conductivity also has a pronounced $\sqrt T$-type dependence. As depicted in Fig.~\ref{fig3}(d), the magnitude of $\delta\sigma_\mathrm{AH}^{\rm N}(T)$, defined as $\Delta\sigma_\mathrm{AH}(T)/\sigma_\mathrm{AH}^0$, is comparable to that of $\delta\rho_\mathrm{AH}^{\rm N}(T)$, varying from 82\% for sample~\#1 to 290\% for sample~\#3 for $T=2$ K. It is interesting to note that the increase in $\delta\sigma_\mathrm{AH}^{\rm N}(T)$ appears to correlate with strengthening disorder: the zero-temperature resistivity $\rho_0$ increases from 8.2 $\rm m\Omega\cdot {cm}$ in sample~\#1 to 28.7 $\rm m\Omega\cdot {cm}$ in sample~\#3 (see Table~\ref{tabI}).

The observation of such large corrections to the AH conductivity is very surprising if one considers the following facts. 1) It has been predicted based on the symmetry considerations that EEI would not lead to any correction to AH conductivity~\cite{Langenfeld1991,Muttalib2007}, and this prediction has been not challenged experimentally or theoretically. 2) Previously observed low temperature corrections to AH conductivity have been attributed to the weak localization~\cite{Mitra2007,Lu2013,Ding2015,Wu2016}, but in our experiment many characteristics of $\sigma_{xx}$ and $\sigma_{xy}^\mathrm{OH}$ strongly suggest that this effect is negligible in $n$-HgCr$_2$Se$_4$. 3) The ratio between the relative corrections to the AH and longitudinal conductivity, namely $r_\sigma=\delta \sigma_\mathrm{AH}^\mathrm{N}(T)/\delta\sigma_{xx}^\mathrm{N}(T)$, has an absolute value over 80 for all three samples (see Table~\ref{tabI}), in stark contrast with much weaker correction observed in any other material, in which $\left|r_\sigma\right|$ is no more than 1.2~\cite{Mitra2007,Lu2013,Ding2015,Wu2016,SM}. 4) In previous experiments on FM metals or magnetic semiconductors, both $\rho_{xx}$ and $|\rho_\mathrm{AH}|$ were found to increase with decreasing temperature~\cite{Mitra2007,Lu2013,Ding2015,Wu2016}, whereas in HgCr$_2$Se$_4$ the increase in $\rho_{xx}$ is accompanied by decreasing $|\rho_\mathrm{AH}|$ as the temperature drops. Therefore, the giant corrections to AH conductivity observed in this work can be neither classified into any known type of FM materials nor described by existing theory.

\begin{figure}[htb]
	\centering
	\includegraphics[width=8.5cm]{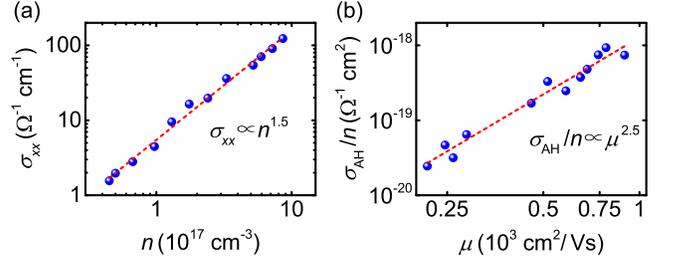}
	\caption{(a) Log-log plot of $\sigma_{xx}$ as a function of carrier density $n$. The data roughly follow a power law $\sigma_{xx}\propto n^{1.5}$.  
		 (b) Scaling plot of $\sigma_\mathrm{AH}/n$ versus carrier mobility $\mu$, where $n$ is the carrier density, the mobility is evaluated with $\mu=\sigma_{xx}/(ne)$. The data in both panels were obtained at $T=2$ K.}
	\label{fig4}
\end{figure}

The nontrivial nature of the AHE in HgCr$_2$Se$_4$ is further revealed in Fig.~\ref{fig4}, which compiles the AHE data acquired at $T=2$ K from many HgCr$_2$Se$_4$ samples.  The carrier density is varied over one order of magnitude and the conductivity roughly follows a power law $\sigma_{xx}\propto n^{1.5}$. 
Following Ref.~\cite{Lee2004}, we plot $\sigma_\mathrm{AH}/n$ as a function of carrier mobility $\mu$ in Fig.~\ref{fig4}(b), which shows the data can be fitted into a scaling law $\sigma_\mathrm{AH}/{n}\propto \mu^{\alpha}\propto \tau^{\alpha}$ with $\alpha\approx 2.5$, where $\tau$ is the transport life time. For the samples of low disorder, one would expect $\alpha=0$ and $\alpha=1$ respectively for the intrinsic/side jump and skew scattering mechanisms~\cite{Nagaosa2010}, since the former would lead to $\sigma_\mathrm{AH}\propto \tau$, and the latter $\sigma_\mathrm{AH}\propto \tau^0$. On the other hand, it is interesting to note that Xiong et al.~\cite{Xiong2010} observed in ultrathin CNi$_3$ thin films that there was a crossover from $\alpha=1.6$ to $\alpha=2$ as the disorder strength became greater than a threshold (corresponding to $k_F l_e < 4$). Their tunneling conductance measurements suggest that the $\alpha=2$ regime is accompanied by a significant suppression of the density of states at the Fermi energy due to the EEI~\cite{Xiong2010}.  Unfortunately, the temperature dependence of the AHE in this system is not given in Ref.~\cite{Xiong2010}, so a direct comparison between the EEI effects in CNi$_3$ and HgCr$_2$Se$_4$ is not possible. In addition to the EEI under strong disorder, the large $\alpha$ value extracted for HgCr$_2$Se$_4$ at $T=2$\,K may also be related to variations in the Berry phase contribution and its low temperature correction among different samples, because of more than one order of magnitude change in the carrier density, as shown in Fig.\,4(a).

The pronounced EEI effects on $\sigma_{xx}$ and the OH conductivity consistent with conventional understanding can be interpreted as being due to a correction to the density of states of the Altshuler-Aronov type~\cite{Altshuler1985, Lee1985}. With the AH transport provided by scatterings on the Fermi surface~\cite{Haldane2004}, it is natural to expect the density of states correction to have a similar effect on $\sigma_\mathrm{AH}$, leading to the $\sqrt{T}$-dependences for the AH conductivity. Further quantitative analysis must await an analytical model of spin-orbit coupling in this material, in which the ferromagnetism and electron transport are contributed separately from a 3D Heisenberg-type network of Cr$^{3+}$ ions and very low-density itinerant electrons. Such a unique electronic structure brings the electron transport into a regime that has not been explored in detail before. It is also noteworthy that previous conclusion of the absence of EEI correction to the AH conductivity was built on the models of skew scattering and side jump and also required presence of the mirror symmetry~\cite{Langenfeld1991,Muttalib2007}. In order to account for the surprising corrections to AH conductivity observed in HgCr$_2$Se$_4$, it may be necessary to consider  the intrinsic terms related to the band Berry curvature ~\cite{Nagaosa2010,Xiao2010}, which were not considered in Refs.\,~\cite{Langenfeld1991,Muttalib2007}. In addition, it may be also helpful to conceive realistic models of impurity effects while taking the magnetic exchange, spin-orbit interaction, and EEI into account~\cite{Culcer2011,Averkiev2002,Arakawa2016}, and this might open a door to broken mirror symmetry~\cite{Note2}.


In summary, we have measured electron transport properties in $n$-HgCr$_2$Se$_4$,  a half-metallic FM semiconductor with extremely low electron densities~\cite{Guan2015,Lin2016}.  This unique electronic structure enabled us to measure the low temperature corrections to longitudinal, OH and AH resistivities in the same samples for the first time. 
For the samples with sufficiently weak disorder, the temperature dependences of longitudinal and OH conductivities can be described by the EEI theory satisfactorily~\cite{Altshuler1985,Lee1985}, whereas the correction to the AH conductivity is more than one order of magnitude larger than those observed in other FM systems~\cite{Mitra2007,Lu2013,Ding2015} and defies previously predicted absence of the EEI correction to the AH conductivity~\cite{Langenfeld1991,Muttalib2007}. Even though our experimental results strongly suggest a link between EEI and the large low temperature correction to AHE, further work, especially from the theoretical side, is highly desirable for resolving the discrepancy between our experiment and the existing theory.

\begin{acknowledgments}
SY and ZL contributed equally to this work. We are grateful to Y. L. Chen and A. J. Liang for sharing their unpublished ARPES data. We thank X. Dai, Z. Fang, H.-Z. Lu, S.-Q. Shen, J. R. Shi, H. M. Weng, H. Y. Xie, and P. Xiong for valuable discussions. This work was supported by the National Key Research and Development Program (Project No.~2016YFA0300600), the National Science Foundation of China (Project No.~61425015), the National Basic Research Program of China (Project No.~2015CB921102), the Strategic Priority Research Program of Chinese Academy of Sciences (Project No.~XDB28000000), and the National Postdoctoral Program for Innovative Talents of China Postdoctoral Science Foundation (Project No.~BX201700012).
\end{acknowledgments}



%

\end{document}